%% file: arXivSubmission.tex
\documentclass[12pt]{article}
\usepackage{amsmath} 
\usepackage{amsthm} 
\usepackage{amssymb} 
\usepackage[colorlinks=true,linkcolor=black,citecolor=blue]{hyperref}
\usepackage{graphicx}
\usepackage[font=small,labelfont=bf]{caption}
\usepackage{filecontents}
\usepackage{subcaption}
\usepackage[table,xcdraw]{xcolor}
\usepackage{tikz}
\usepackage{siunitx}
\usepackage{tabularx} 
\usepackage{lipsum}
\usepackage{booktabs}
\usepackage[margin=1in]{geometry} 
\usepackage[english]{babel}
\usepackage[autostyle, english = american]{csquotes}
\usepackage{siunitx}
\usepackage{float}
\usepackage{array}
\usepackage{paralist}
\usepackage{setspace} 
\MakeOuterQuote{"} 
\usepackage{cite}
\usepackage{multicol} 
\usepackage{wrapfig}
\usepackage{adjustbox}
\usepackage{physics}

\renewcommand{\vec}[1]{\textbf{#1}}

\newcommand{\HWPW}[1][ ]{path weight#1}
\newcommand{\HWPWs}[1][ ]{path weights#1}

\begin{document}
\vspace{3ex}
\begin{center}
\textbf{\large Conserved Quantities in Models of Classical Chaos}
\vspace{3ex}\\
Henry Ando,$^{1,2}$ David A. Huse$^1$
\vspace{1ex}\\
$^1$\textit{Department of Physics, Princeton University, Princeton, New Jersey 08544, USA}
\\
$^2$Currently at: \textit{Department of Physics, University of Chicago, Chicago, Illinois 60637, USA}
\\
(Dated: May 6, 2019. Submitted: July 17, 2023)
\end{center}
\thispagestyle{empty} 

Quantum chaos is a major subject of interest in condensed matter theory, and has recently motivated new questions in the study of classical chaos. In particular, recent studies have uncovered interesting physics in the relationship between chaos and conserved quantities in models of quantum chaos. 
In this paper, we investigate this relationship in two simple models of classical chaos: the infinite-temperature Heisenberg spin chain, and the directed polymer in a random medium. We relate these models by drawing analogies between the energy landscape over which the directed polymer moves and the magnetization of the spin chain. We find that the coupling of the chaos to these conserved quantities results in, among other things, a marked transition from the rough perturbation profiles predicted by analogy to the KPZ equation to smooth, triangular profiles with reduced wandering exponents. These results suggest that diffusive conserved quantities can, in some cases, be the dominant forces shaping the development of chaos in classical systems.

\vspace*{\fill}
\newpage

\begin{titlepage}
\centering
\Large{\textbf {Conserved Quantities in Models of Classical Chaos}}
\vspace{10mm}

\Large\textit{\textbf{Henry A. Ando}} \\
\vspace{25mm}
\large{Junior Paper \\ Department of Physics \\ Princeton University \\Spring 2019\\}
\begin{figure}[H]
\centering
\includegraphics[scale=0.1]{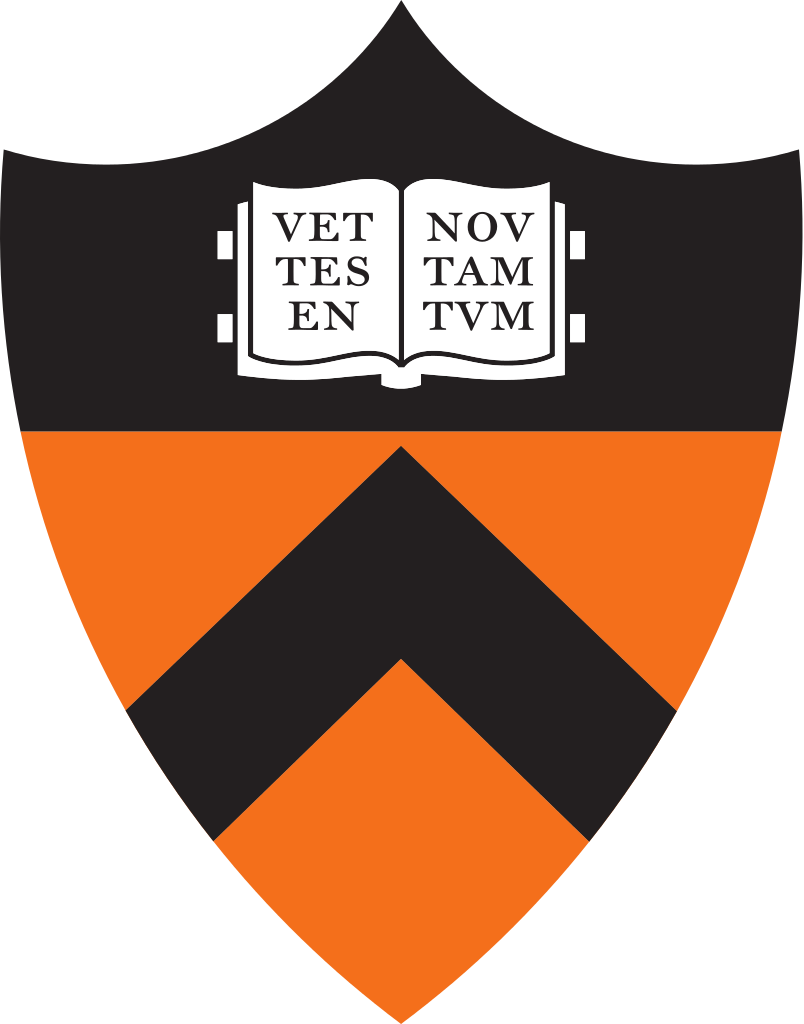}
\end{figure}
\vspace{1mm}
\normalsize{Advised by Professor David A. Huse}\\

\vfill

\end{titlepage}

\newpage

\tableofcontents
\thispagestyle{empty} 

\newpage
\pagenumbering{arabic}

\section{Introduction}

\subsection{Motivation}

Quantum chaos is a major topic of study in condensed matter physics. Although this project centers on classical chaos, it arises out of questions originally asked about quantum systems. In particular, recent work has shown that the introduction of diffusive conserved quantities has important consequences for the evolution of chaotic quantum systems. In \cite{khemani2018operator}, the authors found very interesting physics in a random quantum circuit model with a locally conserved quantity that moves diffusively. Their results help explain how unitary quantum dynamics result in dissipative hydrodynamics on long time scales. In \cite{rakovszky2019sub}, the authors found that R\'eyni entropies $S_{\alpha}$ with $\alpha>1$ exhibit sub-ballistic ($\sqrt{t}$) growth in the presence of diffusive conserved quantities. These results indicate that a better understanding of the interplay between chaos and conserved quantities will be an important step in understanding chaos in more complicated, more physically realistic systems.

In this paper, we study a classical system (the Heisenberg spin chain at high temperatures) with a conserved quantity that moves diffusively (the magnetization). Rather than study a continuous time evolution, we implement a discrete time evolution, which facilitates analogies both to the quantum circuits in \cite{khemani2018operator} and \cite{rakovszky2019sub}, as well as to a very well-studied system, 1+1 directed polymer in random media (DPRM). The DPRM is in some ways simpler than the spin chain model, and thus this analogy will enable clearer insights into the fundamental physics at work in the spin chain. Before describing these models in detail, however, it would be useful to briefly discuss the tools we use to study chaos in general.

\subsection{Lyapunov exponents}
\begin{figure}[t!]
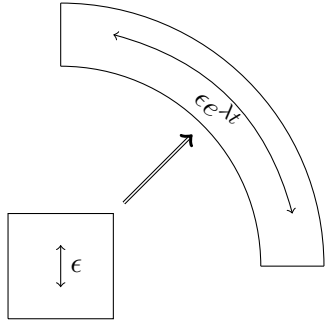

\centering
\include{liouville}
\caption{Diagram showing evolution of a section of phase space in time. Although the total volume is conserved, the section shrinks along one axis and grows along the other axis. Two points separated by a distance $\epsilon$ along the axis with eigenvalue $\lambda$ are separated by a distance $\epsilon e^{\lambda t}$ at a time $t$ later.}
\label{fig:liouville}
\end{figure}

Liouville's theorem states that phase-space volume is conserved under time evolution of a classical Hamiltonian system. In non-chaotic systems, the shape of a section of phase space is roughly preserved over time, and nearby points end up nearby at later times as well. Figure \ref{fig:liouville} illustrates how chaotic systems are different: sections of phase space expand along some axes and contract along others, so even though the total volume is conserved, nearby points end up far away from each other. In general, we characterize this spreading and shrinking with Lyapunov exponents $\lambda_{i}$ \cite{pikovsky2016lyapunov}, which correspond to the rate of stretching or shrinking along the various axes $i$ (one for each dimension of phase space). In the Figure, two points initially separated by a distance $\epsilon$ end up a distance $\epsilon e^{\lambda t}$ apart due to the stretching of phase space along the axis corresponding the exponent $\lambda$. After a while, the behavior of the system becomes dominated by the maximal Lyapunov exponent (MLE). Thus when studying chaos, this is very often a quantity of interest, as it provides a measure of how quickly a small perturbation to the initial conditions blows up.

\subsection{Directed polymers in random media}\label{sec:direct-polym-rand}

At this point, we turn to discussing the models examined in this paper. Although the spin chain model provided the initial motivation, we will introduce the directed polymer model first in the interest of moving from the general to the specific.

Consider a two-dimensional box which is infinitely long in the $x$ direction and has a finite length $T$ in the $t$ direction. Let the bottom of the box be the line $t=0$ and the top of the box be the line $t=T$. A microscopic polymer with some elastic line tension given by\footnote{$\Gamma$ is also analagous to a diffusion constant.} $1/4\Gamma$ will stretch from the origin, at the bottom of the box, to any point at the top of the box. In the zero temperature/high tension limit, the polymer will stretch straight from $(0, 0)$ to $(0, T)$. At finite temperatures, the polymer will not take a completely straight path; instead, let its path be parameterized by $x (t)$. The energy of the path is then 
\begin{equation}
\int_{0}^{T}\frac{1}{4\Gamma}\left (\frac{dx (t)}{dt}\right)^{2}dt.
\end{equation}
Next, we will add the ``random medium'' component of the model. Define a spatial potential $V (x,t)$, where $V$ is independent at each point in space and Gaussian distributed with zero mean and variance $\mathcal{V}^{2}$. Now the energy of a path $x (t)$ is 
\begin{equation}
\int_{0}^{T}\left[\frac{1}{4\Gamma}\left (\frac{dx (t)}{dt}\right)^{2} + V (x (t),t)\right] dt.
\end{equation}
At this point, one might wonder about the probability distribution of final points on the path. From statistical mechanics, we recall that the probability density of a final point $X$ is proportional to the restricted partition function
\begin{equation}\label{eq:1}
Z (X) = \int_{(0,0)}^{(X,T)}\mathcal{D}x (t) \exp \left (-\beta \int_{0}^{T}dt \left[\frac{1}{4\Gamma}\left (\frac{dx (t)}{dt}\right)^{2}+V (x,t)\right]\right),
\end{equation}
where the large integral is over all paths $x (t)$ starting at $(0,0)$ and ending at $(X,T)$ \footnote{This equation is from \cite{kpz}, and slightly adapted to fit my explanation}.

Although it is possible to analytically treat this continuum version of the DPRM \cite{borodin2016directed,kardar1987replica}, the more common tact has been to translate the problem to a discrete lattice and approach it computationally. Consider a lattice in the $x-t$ plane, with nodes at even $x$ for even $t$, and at odd $x$ for odd $t$. Let the lattice have length $T$ in the $t$ direction, and length $L$ in the $x$ direction with periodic boundary conditions in the $x$ direction. We define $V (x,t)$ essentially the same way, independently assigned to each node with zero mean and variance $\mathcal{V}^{2}$. The energy of a path $x_{t}$ is now 
\begin{equation}
\sum_{t=0}^{T} V (x_{t},t).
\end{equation}
The restricted partition function in equation \eqref{eq:1} translates to 
\begin{equation}\label{eq:2}
Z (X) =  \sum_{\textrm{all paths ending at } X} \exp \left[-\beta \sum_{t=0}^{T}V (x_{t},t) \right],
\end{equation}
As it stands, Equation \eqref{eq:2} is still quite difficult to work with. However, as the temperature goes to zero, $\beta\to \infty$, and we can make the approximation that Equation \eqref{eq:2} is dominated by the largest term in the sum. In this case, the polymer simply follows the path of lowest energy through the medium, and so the problem of calculating the statistics of the DPRM turns into a problem of calculating the statistics of lowest energy paths through random energy landscapes. We will define the ``weight'' $W (x,t)$ of this lowest free energy path as 
\begin{equation}\label{eq:4}
W (x,T) = -\sum_{t=0}^{T}\beta V (x_{t},t) = \sum_{t=0}^{T}z (x_{t},t),
\end{equation}
where we have defined $z (x,t)=-\beta V (x_{t},t)$ for convenience. From now on, we will speak of maximizing weights instead of minimizing free energies.\footnote{The physics of the DPRM is unaffected by this change in sign convention.} We will refer to ``path weights'' $W (x,t)$ and the ``energy landscape'' $z (x,t)$, even though $z$ has the opposite sign of the potential landscape $V (x,t)$.

There are two quantities of interest here. As $T$ increases, the rms displacement $\Delta x= x_{T}-x_{0}$ of the path will also grow as $\sqrt{\langle \Delta x^{2}\rangle}  \sim T^{\zeta}$. This ``wandering exponent'' $\zeta$ describes the competition between the random potential which wants to make the polymer bendy in order to seek out optimal paths, and entropic cost of high displacements which drives the system towards pure diffusion. If the polymer moved diffusively, we would find $\zeta = 1/2$. Instead, numerical studies have found that $\zeta = 2/3$ \cite{kpz, huse1985pinning}. The other quantity of interest is the standard deviation of the \HWPWs $W (x,T)$, which also follow a power law $T^{\omega}$. This ``\HWPW fluctuation index'' $\omega$ has been shown to be $1/3$ \cite{kpz}.

This lattice version of the 1+1 DPRM has many applications, including directed percolation \cite{kpz}, tearing of paper sheets \cite{kertesz1992fractal,kertesz1993self}, domain boundaries in dirty Ising magnets \cite{huse1985pinning}, blown fuse networks \cite{hansen1991roughness}, and, as we will see in the next section, chaos in lattices.

\subsection{One dimensional discrete-time lattices}\label{sec:one-dimens-discr}

Now we will turn from studying the statistical mechanics DPRM problem to a deterministic dynamical model for a general, discrete-time chaotic system. In this situation, the averages will be ensemble averages over all possible initial states, rather than the statistical mechanics DRPM where averages are viewed in a slightly different light. However, the physics (as we will show in this section) are essentially the same.

Consider a $1+1$ dimensional lattice with the same geometry as the previous section (diamond-shaped, with nodes on even $x$ positions at even $t$, and odd $x$ positions at odd $t$). Let each node have a parameter $s_{x,t}$. Suppose that $s_{x,t}$ is determined by some function $f (s_{x-1,t-1}, s_{x+t,t-1})$ of the two sites before it, so the entire grid is determined by the values at $t=0$. 

What we would like to study is how small perturbations to a site $s_{a,0}$ influence a site $s_{x,t}$. If the dynamics of grid are chaotic, this would be an excellent way of characterizing the chaos, as it would enable us to compute the exponential-in-time Lyapunov growth of the perturbation. To compute $ds_{x,t}/ds_{a,t}$, we need to use the chain rule along all paths starting at $(a,0)$ and ending at $(x, t)$, so 
\begin{equation}
\frac{ds_{x,t}}{dx_{a,0}} = \sum_{\textrm{all paths } x_{t}} \frac{ds_{x_{t},t}}{ds_{x_{t-1},t-1}}\cdot \frac{ds_{x_{t-1},t-1}}{ds_{x_{t-2},t-2}}\cdots \frac{ds_{x_{1},1}}{ds_{x_{0},0}},
\end{equation}
where each path $x_{t}$ has endpoints $x_{0}=a$ and $x_{t}=x$. 

This seems fairly complicated, but in a ``low-temperature'' limit where one of the terms in the sum is much bigger than the others, it reduces to a single term which is the product of step-wise derivatives along the highest-weight path 
\begin{equation}
\frac{ds_{x,t}}{dx_{a,0}} = \frac{ds_{x_{t},t}}{ds_{x_{t-1},t-1}}\cdot \frac{ds_{x_{t-1},t-1}}{ds_{x_{t-2},t-2}}\cdots \frac{ds_{x_{1},1}}{ds_{x_{0},0}},
\end{equation}
where $x_{t}$ is the highest weight path. Taking the log of both sides reduces the product into a sum, 
\begin{equation}
\log\left (\frac{ds_{x,t}}{dx_{a,0}}\right) = \log\left (\frac{ds_{x_{t},t}}{ds_{x_{t-1},t-1}}\right) + \log\left (\frac{ds_{x_{t-1},t-1}}{ds_{x_{t-2},t-2}}\right) + \cdots + \log \left (\frac{ds_{x_{1},1}}{ds_{x_{0},0}}\right).
\end{equation}
We have thus reduced the problem of calculating $ds_{x,t}/ds_{a,t}$ into the problem of finding the highest-weight path from $(a,0)$ to $(x,t)$, where the individual weights are $\log (ds_{x,t}/ds_{x\pm 1,t-1})$. The analogy to the 1+1 lattice DPRM should be clear; with a slight shift in definitions (moving the chain rule from the edges to the nodes), $z(x,t)$ is totally analagous to $\log (ds_{x,t}/ds_{x\pm 1,t-1})$, and we find that this is exactly the same as the DPRM if we make the same assumptions on the distribution of $\log (ds_{x,t}/ds_{x\pm 1,t-1})$ (that it is Gaussian).

\subsection{The classical Heisenberg spin chain}\label{sec:class-heis-spin}

\begin{figure}[t!]
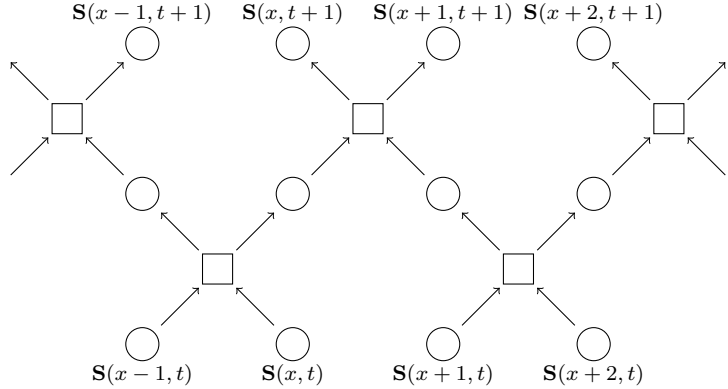

\centering
\include{interactionscheme}
\caption{Diagram of the time evolution of the system. In one time step, the spins each interact with one of their neigbors, then the other. The pairwise spin interactions (squares in the figure) given in Equation \eqref{eq:20}. }
\label{fig:InteractionScheme}
\end{figure}

We turn now to the model system which provided the initial motivation for this project: the classical Heisenberg spin chain, with nearest-neighbor interactions. The Hamiltonian of this system is 
\begin{equation}\label{eq:3}
H = -J\sum_{x=0}^{L-1}\vec{S} (x)\cdot \vec{S} (x+1),
\end{equation}
where the $\vec{S}$ are unit-length 3-component vector spins, $J>0$, and we have assumed periodic boundary conditions $\vec{S} (x)\equiv \vec{S} (x+L)$. The dynamics of this Hamiltonian conserve both total energy and total magnetization. To simplify the numerics, we use here a discrete-time dynamics with a large time step. We will design the precessional dynamics so as to conserve the magnetization but not the energy. The basic scheme of these discrete-time dynamics are shown in Figure \ref{fig:InteractionScheme}. Each spin interacts with one of its neighbors in the first half of a time step, and with its other neighbor in the second half of a time step. In the pairwise spin interactions (squares in the Figure), each spin precesses around the axis defined by the sum of the two spins, by an angle determined by the magnitude of this sum. In math, this means that 
\begin{equation}\label{eq:20}
\vec{S} (x, t+\frac{1}{2}) =  \vec{R} (\vec{n}, \theta) \vec{S} (x, t),
\end{equation}
where $\vec{R} (\vec{n}, \theta)$ is the matrix encoding a rotation around the vector $\vec{n}$ by an angle $\theta$, and where 
\begin{equation}\label{eq:21}
\vec{n}= \frac{\vec{S} (x, t) +\vec{S} (x\pm 1, t)} {|\vec{S} (x, t)+\vec{S} (x\pm 1,t)|},
\end{equation}
and 
\begin{equation}\label{eq:12}
\theta=\eta (|\vec{S} (x, t)+\vec{S} (x \pm 1, t)|)
\end{equation}
for some function $\eta (x)$. We initially used $\eta (x)=A x$ for a constant $A=\pi/4$,\footnote{This value was chosen so that the largest possible rotation angle would be $\pi/2$.} but eventually tried more complicated functions to try to make the magnetization more diffusive. 

These dynamics conserve the total magnetization of the system and preserve phase space volume (Liouville's theorem). Energy is not conserved, but for the purposes of this investigation energy conservation is unnecessary. We will investigate the coupling of the Lyapunov exponents to the magnetization, largely ignoring the energy. When the system is strongly magnetized, the chaos will be weaker than when the system is unmagnetized, and this is the behavior that we will explore.

To study chaos in this model, we consider a perturbed initial state, $\vec{S}' (x, 0)=\vec{S} (x, 0)+\epsilon \vec{P} (x, 0)$, with $\vec{P} $ perpendicular to $\vec{S} $ and $\epsilon$ small. We then evolve the system as before, keeping only the terms that are first order in $\epsilon$.
As we will show, the magnitude of the perturbation grows exponentially under these dynamics. The mean exponent will be the maximal Lyapunov exponent $\lambda$ of this system. 

The analogy to the very general discrete-time one-dimensional lattice problem discussed in Section \ref{sec:one-dimens-discr} should hopefully be clear. We have a lattice with parameters $\vec{S} (x,t)$, which interact pairwise in discrete time steps, and we want to calculate the derivative of $\vec{S} (x, t)$ with respect to changes in the initial conditions $\vec{S} (x', 0)$. By tracking the perturbation $\vec{P}$ to first order, we do exactly this. In this analogy, $\log (|\vec{P} (x, t)|)$ is the \HWPW $W (x, t)$ of the directed polymer terminating at $x$, and the pairwise interactions between spins form the energy landscape $z (x, t)$ through which the directed polymer moves. Unlike the DPRM, we have no parameter (like temperature) that would allow us to go to a limit where the sum is dominated by one highest-weight path. This model will therefore allow us to examine a slightly more complicated version of what we will explore in the DPRM.

This system, however, has an additional feature not present in the DPRM discussed in Section \ref{sec:direct-polym-rand}. Here, the magnetization is conserved and moves diffusively, and if the local ``energies'' $z (x,t)$ are indeed coupled to the magnetization as we will later show, the spin chain is analagous to a directed polymer on an energy landscape where the energies are \textit{not} independent at each site, but rather move diffusively. We will discuss this modification to the directed polymer in the next section.

\subsection{Energy conservation in the DPRM}

In a normal simulation of the 1+1 diamond lattice DPRM like the one discussed in Section \ref{sec:direct-polym-rand}, each site $(x,t)$ is assigned a random Gaussian distributed variable $z (x,t)$, with zero mean and unit variance. The weights $W (x,t)$ of the highest weight paths through this medium are then calculated via the following simple recursive algorithm: 
\begin{equation}\label{eq:6}
W (x, t+1)=\max \left[ W (x-1, t)+z (x-1,t), W (x+1,t)+z (x+1, t)\right].
\end{equation}
We can also keep track of the paths themselves. 

We now introduce a modified medium where we assume that the total energy $\sum z (x,t)$ are conserved. This is done as an analogy to the spin chain, where the total magnetization $\sum \vec{S} (x,t)$ is conserved. We will show later that high magnetization behaves similarly to low $z$ in the DPRM. Thus a DPRM where $z$ moves diffusively will be similar to the spin chain, where the magnetization moves diffusively. 

To construct this modified medium, the initial energies $z (x,0)$ are selected (as before) from a Gaussian distribution with zero mean and unit variance. The energies at later times are then determined by 
\begin{equation}\label{eq:5}
z (x, t+1) = \frac{1}{2}\left[ z (x-1, t) + z (x+1, t) + j (x-1, t) - j (x+1,t) \right],
\end{equation}
where the $j (x,t)$ are random currents (also Gaussian distributed with zero mean and unit variance) which introduce some randomness by laterally moving energy between neighboring sites. Under this process, there are no same-time spatial correlations. As $t_{0}$ becomes large, $\langle z(x+x_{0},t+t_{0}) z (x,t)\rangle $ approaches a Gaussian distribution in $x_{0}$, whose standard deviation scales like $\sqrt{t_{0}}$.

\begin{figure}[b!]
\centering
\includegraphics{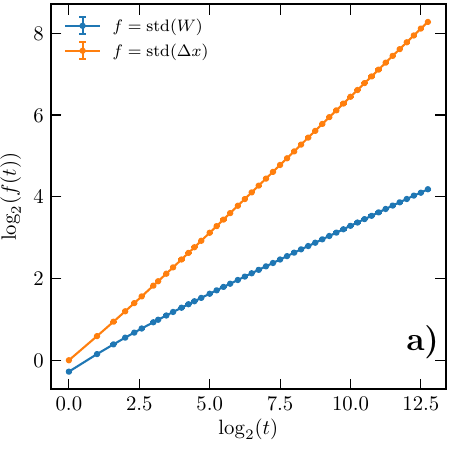} \hfill
\includegraphics{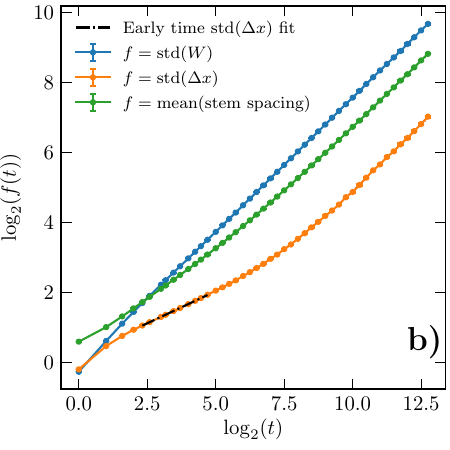}
\caption{Log-log plots of various DPRM characteristics versus time, for a) the standard DPRM with independent Gaussian weights on each site, and b) the doubly-highest weight paths of the DPRM with energy dipole conservation. Each figure represents the average of 128 simulations, each with $T= 8192$ and $L= 32768$. Error bars (present, although not visible) represent the standard error of the mean on each point. The fitted parameters in Table \ref{tab:params} come from fitting to the data from $t=512$ and beyond. Subplot b) includes a fit to the $\textrm{std} (\Delta x)$ at early times ($4<t<32$), which yielded a slope of $\zeta_{1}' = 0.367 \pm 0.001$, in contrast to the later time slope $\zeta_{1} = 0.758 \pm 0.052$. The rest of the scaling figures have been left out because mostly, they all look like subplot a) but with different slopes. Subplot b) was included because of its unique curvature.}
\label{fig:scaling}
\end{figure}

\section{Results}

\subsection{DPRM}

\subsubsection{Groundwork: the DPRM without energy conservation}

We begin by presenting some characteristics of the DPRM with uncorrelated energies $z (x,t)$, which we will then compare to the DPRM with energy conservation.
First, to confirm that our simulations and data analysis techniques were working properly, we reproduced the literature result that $\omega=1/3$ and $\zeta = 2/3$ \cite{huse1985pinning,kpz}. We performed $N$ normal DPRM simulations with total time $T$ and total length $L$, with periodic boundary conditions. For each time step $t$ of each simulation, we kept track of the standard deviation over all positions $x$ of $W (x,t)$, and of $\Delta x (x,t)$, which is defined to be the displacement between $x$ and the starting point of the path that led to $(x, t)$. Then we averaged these values over all iterations, keeping track of the standard errors, and plotted them on a log-log plot such as Figure \ref{fig:scaling}a. To extract the slopes of these lines and thus the scaling exponents of the parameters, linear fits to the last half of the log-log plot were performed in Python (the first 512 data points were neglected to focus on the leading-order scaling powers). Finding the error bars on the fitted slopes using the built in functionality of the Python fitting algorithm would be a mistake, as the program assumes that the errors at each point are uncorrelated - however, the errors on consecutive time steps are highly correlated. Therefore, we found error bars by defining the error on a fitted slope to be the difference between the slopes of the flattest and steepest lines that would fit within the error bars of the fitted region. 
Table \ref{tab:params} shows that our values for $\omega$ and $\zeta$ do reproduce the literature values of $1/3$ and $2/3$ to within the error bars.

\begin{table}[t!]
\centering
\begin{tabular}{| c || c  | c | c |}
\hline
Parameter & No conservation & 
Energy conservation & 
Energy dipole conservation \\\hline

$\omega$ & $0.328 \pm 0.008$ & $0.687 \pm 0.005$ & $0.769 \pm 0.006$ \\

$\zeta$  & $0.666 \pm 0.003$ & $0.708 \pm 0.004$ & $0.765 \pm 0.005$ \\

$\omega_{1}$ & $0.327 \pm 0.010$ & $0.687 \pm 0.005$ & $0.768 \pm 0.006$ \\

$\zeta_{1}$ & $0.661 \pm 0.006$ & $0.625 \pm 0.010$ & $0.758 \pm 0.010$ \\

$\rho$ & $0.667 \pm 0.003$ & $0.701 \pm 0.004$ & $0.752 \pm 0.004$ \\ \hline

\end{tabular}
\caption{Table of fitted DPRM parameters. $\omega$ and $\zeta$ apply to all highest weight paths, and $\omega_{1}$ and $\zeta_{1}$ apply to the doubly highest weight paths. $\rho$ is the scaling power of the mean spacing between stems. Fitted values are the slopes of linear fits to plots like Figure \ref{fig:scaling}, which represent the average of 128 simulations with $T=8192$ and $L=32768$. Error bars are given by the formula in Equation 8. }
\label{tab:params}
\end{table}

There are a few other things to examine in the DPRM before introducing energy conservation. One is the actual geometry of the highest weight paths themselves. Figure \ref{fig:paths}a shows the highest weight paths ending at each point $(x,T)$ of a DPRM simulation with $T=1000$. These path networks are scale-free in that zooming in on a region near the top of the plot will yield an image which is visually similar to the plot as a whole. The spatial profile of the path weights, shown in Figure \ref{fig:scalefreedpm}a, is also scale-free. Zooming in on a region, even a steeply sloped one like the one shown in the inset, yields a similarly rough profile. 

The final characteristic to examine before introducing energy conservation is the actual shape of the distribution of \HWPWs $W (x,t)$. It is known that this is an example of the Tracy-Widom distribution \cite{chu2016probability}. Figure \ref{fig:wdistdpm}a shows that, after rescaling the path weights at each time to have zero mean and unit variance, the distributions at different times collapse onto one another.

\subsubsection{Adding energy conservation}

Next we introduce energy conservation to the random energy landscape $z (x,t)$, according to the scheme in Equation \eqref{eq:5}. This results in an energy landscape where the energies move diffusively. We therefore expected that the \HWPW fluctuations would grow faster than in the nonconserved case, as polymers in a region of higher or lower energy will stay there for a long time, rather than a single time step. We also expected that the wandering exponent $\zeta$ would decrease, as polymers that found a region of high $z$ would stay there as the region spreads out diffusive ($|\Delta x|\sim t^{1/2}$), which is slower than the original wandering ($|\Delta x|\sim t^{2/3}$).

\begin{figure}[h!]
\centering
\includegraphics[scale=0.95]{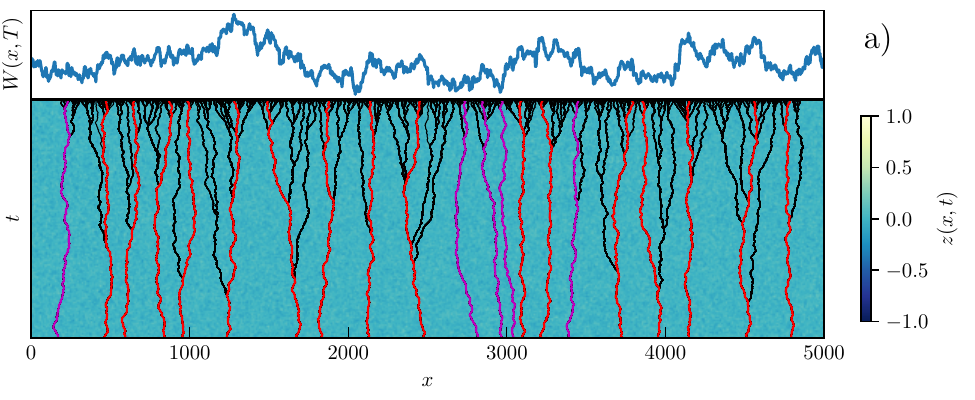}
\includegraphics[scale=0.95]{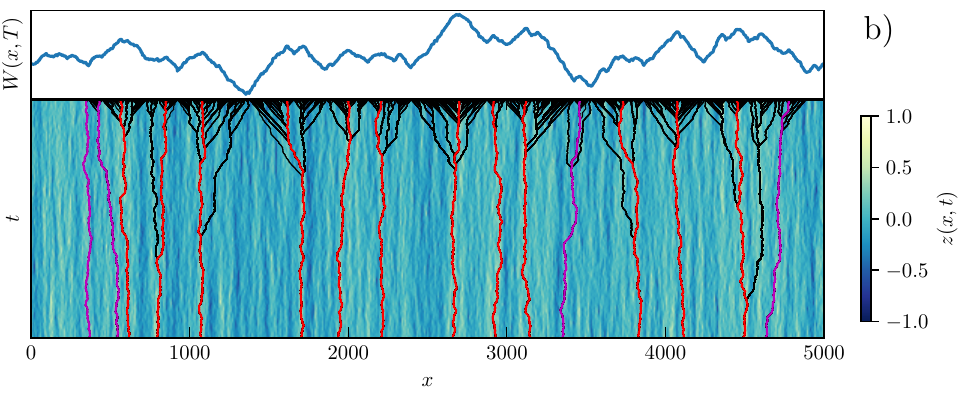}
\includegraphics[scale=0.95]{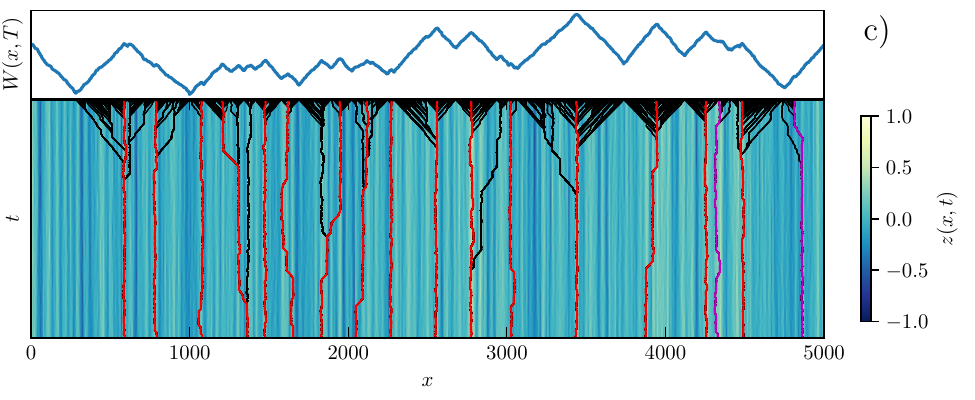}
\caption{Highest weight path for the DPRM with a) no energy conservation, b) energy conservation, and c) energy dipole conservation. The energy landscape $z (x,t)$ is plotted as a colormap in the background of the path diagram, with Gaussian smoothing of radius 4 in $x$ and $t$. The weight $W (x,T)$ at the final time $T$ is plotted above each path diagram. Doubly highest weight paths are plotted in red, and occluded doubly highest weight paths are plotted in magenta. Each simulation has length $L=5000$ and total time $T=1000$. }
\label{fig:paths}
\end{figure}

Table \ref{tab:params} shows that we were correct that the energy fluctuations would grow faster: $\omega= 0.687$ in this modified DPRM. However, we found that the wandering exponent $\zeta$ actually increased to $0.708$, counter to what we expected. This result can be explained by examining Figure \ref{fig:paths}b. The polymers through this diffusive energy landscape are qualitiatively quite different from the uncorrelated landscape. Now, the path networks resemble a diagonally branching ``flower'' on top of a long, nearly straight ``stem'' with relatively few branching points. These networks are no longer scale-free; there is one region which consists of just stems, and another region near the top which consists of just flowers. This separation into two very different regimes explains why the wandering exponent failed to decrease: while the stems do appear to wander less, most of the paths do not stick to the stem, but ultimately branch out into a flower with a much greater width. Thus the wandering is dominated by the spacing between stems, not the wandering within them.

\begin{figure}[t!]
\centering
\includegraphics{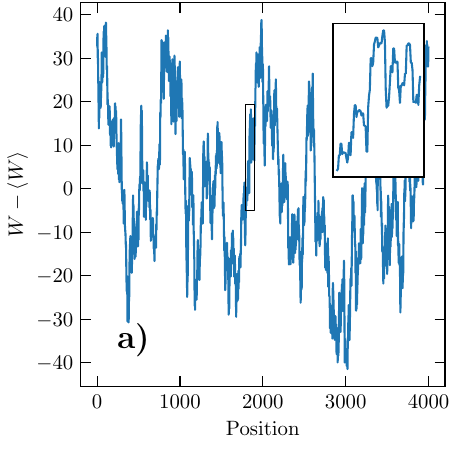}\hfill
\includegraphics{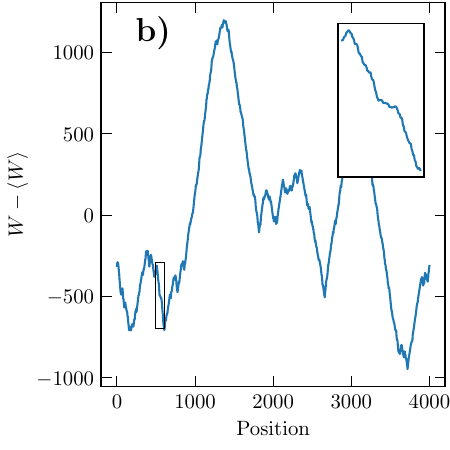}
\caption{Plots of the path weight $W (x,t)$ vs. $x$ in the DPRM, with a) no energy conservation and b) energy conservation. Insets show the contents of the small boxed region of the plot. For these images, $t=1024$.}
\label{fig:scalefreedpm}
\end{figure}

To try to separate these two effects, we also found the highest weight path within each flower, and recorded the wandering and the energy fluctuations of these ``doubly-highest weight'' paths. We found that the doubly-highest weight path wandering exponent $\zeta_{1}$ was, in fact, lower, at around 0.625. Moreover, we found that the spacing between stems grows as a power law $t^{\rho}$, and that this ``stem spacing exponent'' $\rho$ is 0.701, which is almost within error bars of the wandering exponent $\zeta$. 

In the DPRM without energy conservation, the doubly-highest weight path exponents $\omega_{1}$ and $\zeta_{1}$ are within error bars of $\omega$ and $\zeta$, which emphasizes the fact that there are not two separate regimes in this case. Indeed, the stem spacing exponent $\rho =2/3$ as well.

There is another strong qualitative difference, this time in the spatial profile of the \HWPWs[]. Figure \ref{fig:scalefreedpm}b shows that this plot, which was scale-free without energy conservation, now clearly has scale. The zoomed-in inset shows none of the roughness that previously characterized this profile. Indeed, looking back at Figure \ref{fig:paths}b, it seems clear what is happening: a stem path grows through a region of relatively high $z$ and accumulates a \HWPW which is substantially greater than the \HWPWs of all other nearby paths. For all the points in the flower of this stem, the highest weight path is to move as directly 
as possible (diagonally) to the stem, then follow it all the way down. In the approximation that the energies $z (x,t)$ in between the stem and the path source are all zero, the spatial profile of the \HWPWs will be a triangle, which is exactly what appears to be happening in the 
\newpage
\clearpage
\newpage
\clearpage
\noindent
top part of Figure \ref{fig:paths}b.

A third difference is in the distribution of \HWPWs[]. Figure \ref{fig:wdistdpm}b shows that while the shape of the distribution is roughly the same, the amount of skew appears to be increasing over time.

At this point, we wanted to see whether we could exaggerate the above effects to make it even clearer what was going on. To this end, we introduced another model, where both the local energy $z (x,t)$ and its dipole moment moment were conserved. This leads to subdiffusive ($t^{1/4}$) transport of the energy.

\subsubsection{Energy dipole conservation}

\begin{figure}[ht]
\centering
\includegraphics[scale=0.5]{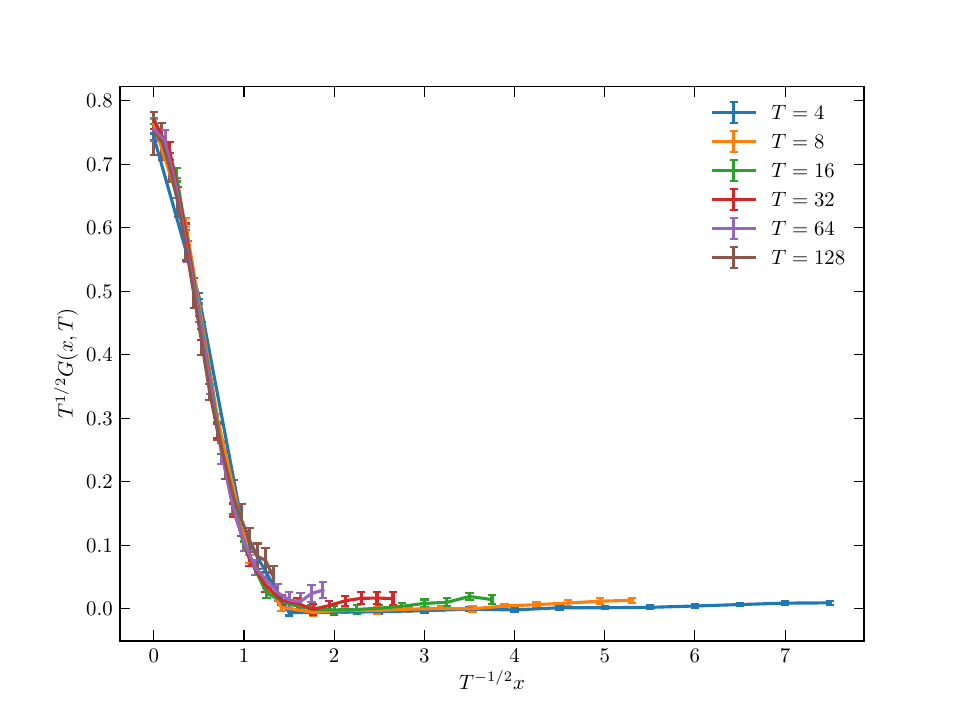}
\includegraphics[scale=0.5]{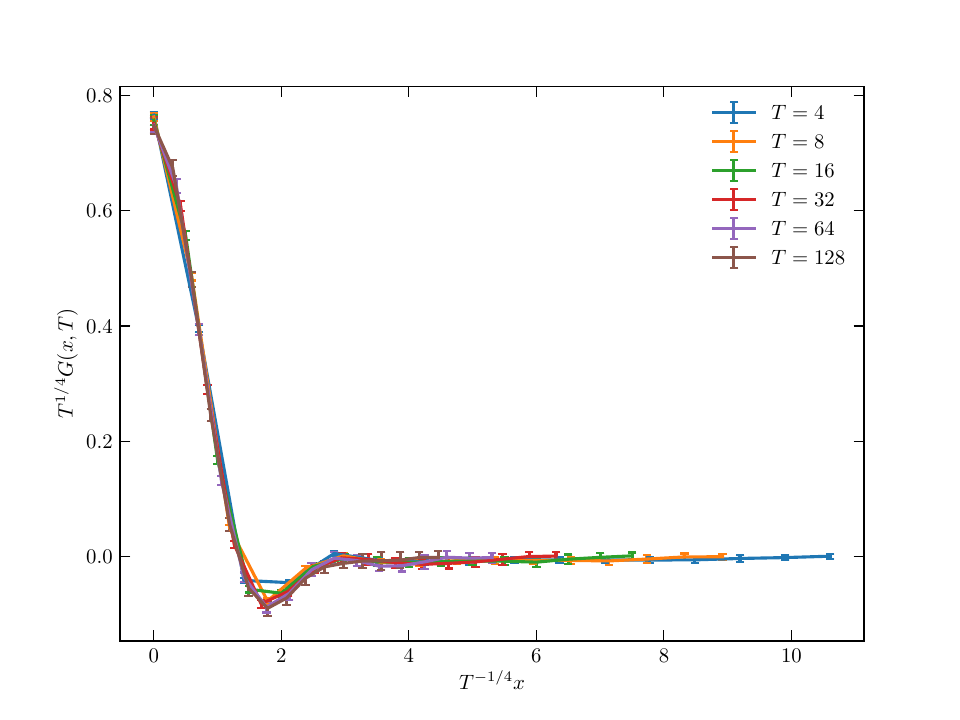}
\caption{Plots of the correlation function $G (x,t)=\langle z (x_{0},t_{0})z (x_{0}+x,t_{0}+t)\rangle$ for different values of $t$ in the DPRM with energy conservation on the left, and the DPRM with energy dipole conservation on the right. Rescaling by $t^{1/2}$ in the left plot and $t^{1/4}$ in the right plot collapses the lines onto one another, showing that the systems exhibit the proper scaling behavior.}
\label{fig:verifyingcorrelation}
\end{figure}

To conserve the energy dipole moment, we replaced Equation \eqref{eq:5} with a modified version,
\begin{align}\label{eq:7}
z (x,t+1) = &\big[ 
-z (x-3,t) + j (x-3, t) + \sqrt{12} j (x-2,t) + 9z (x-1,t)\nonumber 
\\ &-3j (x-1,t) -\sqrt{48}j (x,t) +9 z (x+1, t) + 3j (x+t, t) \nonumber \\
& + \sqrt{12}j (x+2,t) - z (x+3,t) -j (x+3,t)\big] /16,
\end{align}
where $j (x,t)$ are once again random currents with zero mean and unit variance. These dynamics lead, as desired, to $t^{1/4}$ transport of the energy (see Figure \ref{fig:verifyingcorrelation}). We expected that the \HWPW fluctuation index $\omega$ would continue to increase, and that the doubly-highest-weight path wandering exponent $\zeta_{1}$ would continue to decrease.

What we found was surprising. While the \HWPW fluctuation index $\omega$ did increase to $0.769$, confirming one of our predictions, the wandering exponent of the DHWPs actually increased to higher than it originally was. This result is further complicated by the curvature visible in Figure \ref{fig:scaling}b, which shows the wandering and \HWPW fluctuations of 
the DHWPs as a function of time. For all other situations (with all three levels of energy conservation, and for both singly- and doubly-highest-weight paths), the lines have been fairly straight, as in Figure \ref{fig:scaling}a. However, the orange line in Figure \ref{fig:scaling}b seems to have two regimes with completely different slopes, getting much steeper at later times.

Figure \ref{fig:paths}c helps explain what is happening here. Although the stems in this case look overall even straighter than the stems in Figure \ref{fig:paths}b, there are abrupt jumps in the stems at, for example, $x=1300$, $x=1800$, and $x=2000$. Examining the energy landscape $z (x,t)$ underneath, it becomes apparent that the stem has jumped from one ``stream'' of high weight to another.

When does a stem jump from one stream to another? If we approximate that the region between streams has $\langle z (x,t)\rangle = 0$, then the stem will on average pick up no weight while in between the streams. Thus the jump will occur only when the \HWPW difference between being in one stream over the other is greater than the distance between them. But in this dipole-conserved energy landscape, the \HWPW fluctuation index $\omega =0.769$ 
is greater than the stem spacing exponent $\rho = 0.752$, so at long times we expect this to happen frequently - and indeed, at long times, the wandering exponent of the DHWPs $\zeta_{1}$ gets to within error bounds of $\omega$ and $\rho$.

Why doesn't this behavior occur in the previous models? For both, $\omega$ was less than $\rho$, and so jumping events would not have been the dominant source of DHWP wandering.

\begin{wrapfigure}{l}{0.45\textwidth}
\begin{center}
\includegraphics[height=2.25in]{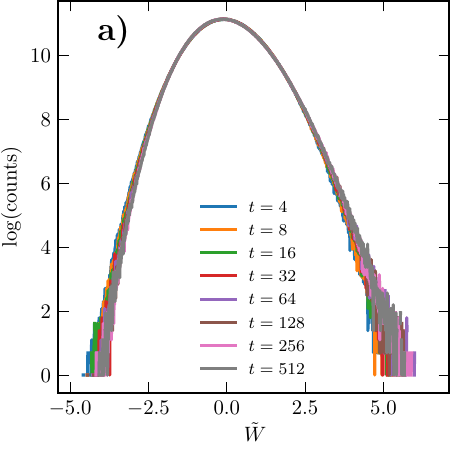}
\includegraphics[height=2.25in]{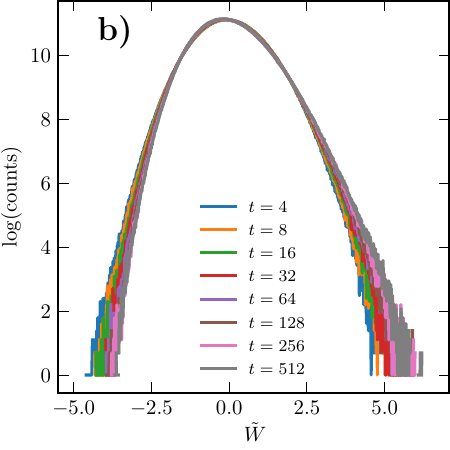}
\includegraphics[height=2.25in]{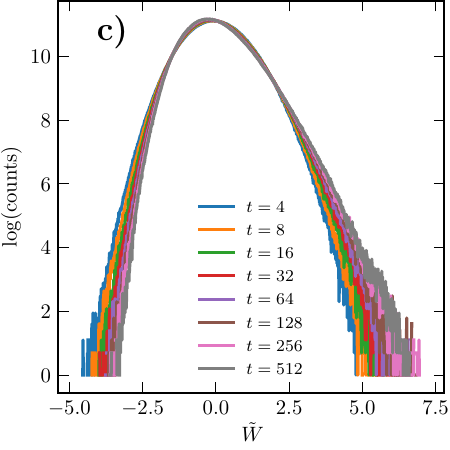}
\end{center}
\caption{Distribution of \HWPWs $W (x,t)$ for the DPRM with a) no energy conservation, b) energy conservation, and c) energy dipole conservation. $\tilde W$ is rescaled to have zero mean and unit variance, so these plots show only the shape and skew of the distributions. Distributions are accumulated from 256 simulations with length $L=1024$.}
\label{fig:wdistdpm}
\end{wrapfigure}

\begin{figure}[b!]
\centering
\includegraphics[width=0.5\textwidth]{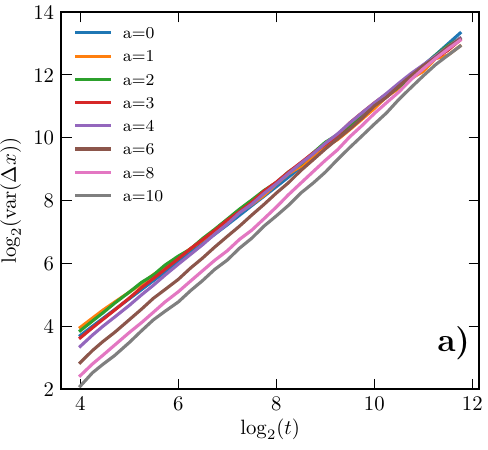}%
\includegraphics[width=0.5\textwidth]{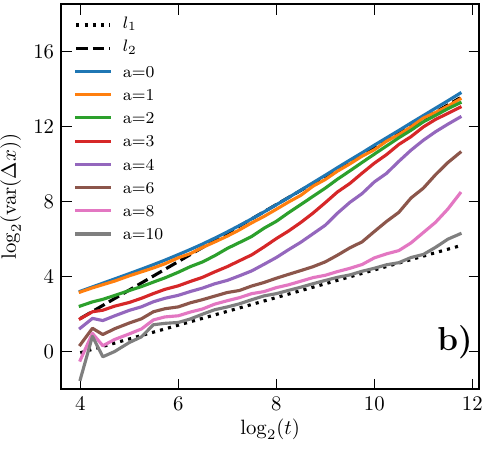}
\caption{Wandering variance of DHWPs beginning at a seed point versus time with a) energy conservation and b) dipole conservation. For these simulations, the initial energies $z (x,0)$ were set to zero rather than a random number - all except the origin, where $z (0, 0)$ was set to a constant $-a$. The DHWPs for each time step were then found, and if any began at $(0,0)$, their wandering was recorded. The $y$ axis of these plots shows this wandering variance, for many different values of $a$. The lines $l_{1}$ and $l_{2}$ have slopes equal to the early and late time wandering exponents of the dipole DHWPs. These plots represent the result of $512$ simulations with $T=4096$ and $L=1024$. }
\label{fig:singleorigin}
\end{figure}

Further evidence of this conclusion is given in Figure \ref{fig:singleorigin}b. For this figure, the initial energies $z (x,0)$ were all set to 0 except for the origin which was set to a constant $-a$. This ``seed'' energy at the origin creates a stream centered there. Over time, other streams form and the weight of this initial stream decreases, and eventually jumping events start to occur. Higher values of $a$ create stronger initial streams, which suppresses the jumping for longer amounts of time. Two straight lines $l_{1}$ and $l_{2}$ are also plotted in Figure \ref{fig:singleorigin}b. These lines have slopes $\gamma_{1}'=0.367$ and $\gamma_{1}=0.758$, which are the early and late-time wandering exponents of the 
dipole-conserved DHWPs from Figure \ref{fig:scaling}b. Qualitatively, it is quite clear that for high values of $a$ the plot follows the slope of $l_{1}$ for a long time, then undergoes a phase transition and converges towards $l_{2}$ as the jumping takes over.

By the other metrics we used to compare the DPRM with and without energy conservation, the dipole-conserved DPRM behaves as expected. Figure \ref{fig:paths}c looks like an exaggerated version of Figure \ref{fig:paths}b, with very straight stems and sharply branching flowers, which result in a very neat, triangular spatial \HWPW profile. And Figure \ref{fig:wdistdpm}c shows that the \HWPW distributions are once again roughly the same shape, with the skew becoming even greater over time than before.

The dipole-conserved DPRM was thus, in some ways, not particularly helpful in understanding the monopole-conserved DPRM. Most of the things it showed were exactly as expected. The jumping transition, while interesting, is not directly useful in understanding the original model. However, it was valuable to confirm that our understanding was correct, and the jumping transition is an interesting enough bit of physics that it seems worth looking further into. Fractons are a current topic of interesting in condensed matter physics, and fracton models often feature precisely these features (the conservation of both a quantity and its dipole moment). It would therefore be interesting to follow up on this phenomenon with a classical chaos model exhibiting sub-diffusive transport of a conserved quantity coupled to the Lyapunov exponent. 

For the moment, however, we will turn to the model which originally inspired this project, the Heisenberg spin chain, and use our findings with the DPRM to guide this investigation.

\subsection{Spin chain}

With the spin chain, an important feature for the present study is the coupling of the Lyapunov exponent to the magnetization. To study this, we first need to set up magnetized initial states. To introduce randomness and magnetization in the initial state, we will create an initial state as if a magnetic field $h$ is applied in the $-z$ direction. Assuming we are at temperature $\beta=1$, and for the purpose of setting this ensemble of initial states ignoring any spin-spin interaction, the probability distribution of the $z$ component of the spin is thus given by the Boltzmann distribution
\begin{equation}\label{eq:8}
P (S_{z}=z) = \frac{1}{Z}e^{hz},
\end{equation}
where $Z$ is the partition function,
\begin{equation}\label{eq:9}
Z = \int_{-1}^{1}e^{hz}dz.
\end{equation}
We sample the $z$ component of each initial spin from this distribution, and then pick the azimuthal angle of the spins at random. The initial perturbations are then chosen to be random unit vectors in the tangent space of the initial spins, $\vec{P}\cdot \vec{S}=0$. The spins are evolved in time as described in Section \ref{sec:class-heis-spin}, and the perturbations are propagated to leading order as described in Section \ref{sec:prop-pert}. Components of the perturbation parallel to the spins should not exist but are an unavoidable consequence of roundoff error, so to prevent roundoff error from propagating, the parallel components of the perturbations are subtracted off after each time step.

\begin{figure}[ht]
\centering
\includegraphics{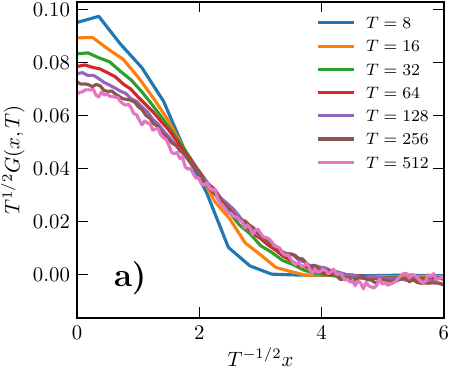}\hfill
\includegraphics{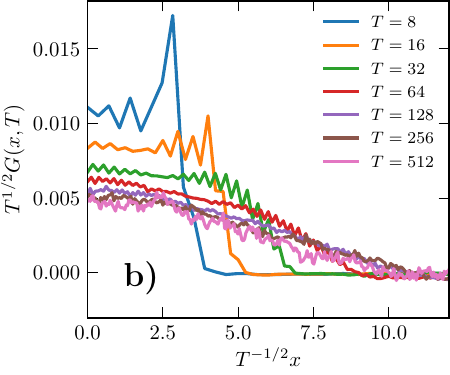}
\includegraphics{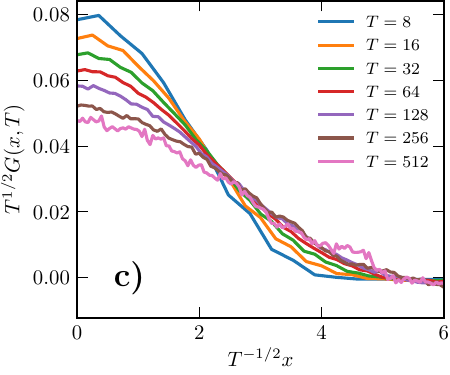}\hfill
\includegraphics{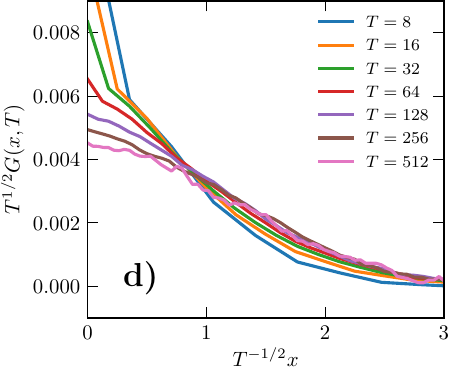}
\caption{Plots of the spin-spin correlation function $G (x,t) $. Subplots a) and b) use $\eta (x) = \pi x /4$ (see Equation \eqref{eq:12}). Subplot a) is at $h=0$, subplot b) is at $h=3.5$. Subplots c) and d) use $\eta_{1} (x) = \pi x (x^{2}-4)^{2}/16$. Subplot c) is at $h=0$, subplot d) is at $h=8$. Each subplot represents the average of 128 simulations with $L=2048$ and $T=1024$.  }
\label{fig:diffusion}
\end{figure}

The next thing we did after checking that the model was functioning properly (perturbations were remaining perpendicular, spins were remaining magnitude 1) was to check whether the magnetization was diffusing. We define the $S_{z}$ correlation function 
\begin{equation}\label{eq:10}
G (x,t) = \big\langle S_{z} (x_{0},t_{0})\cdot  S_{z} (x+x_{0},t+t_{0}) \big\rangle .
\end{equation}
If $S_{z}$ is moving diffusively, we expect that 
\begin{equation}\label{eq:11}
G (x,t) \sim \frac{1}{\sqrt{t}} \exp \left ( -\frac{x^{2}}{4D t} \right)
\end{equation}
for some diffusion constant $D$. Figure \ref{fig:diffusion}a shows that when $h=0$ (no net magnetization), the $z$ component of the spin does move approximately diffusively; when rescaled by $t^{1/2}$, the lines roughly collapse onto one another. However, when $h=3.5$ (Figure \ref{fig:diffusion}b), the behavior is much messier, a subject to which we will return later. 

\begin{figure}[b!]
\centering
\includegraphics[]{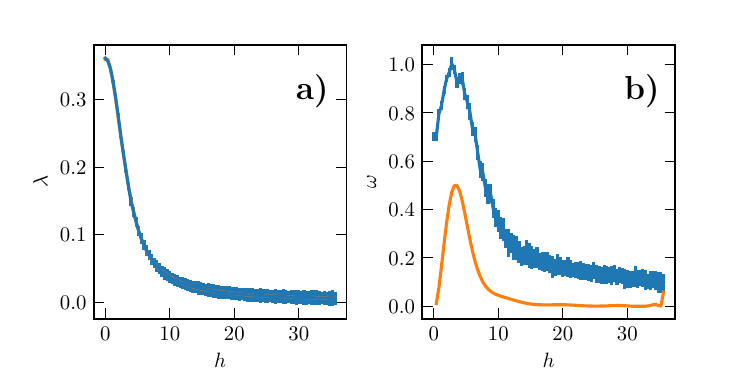}
\includegraphics{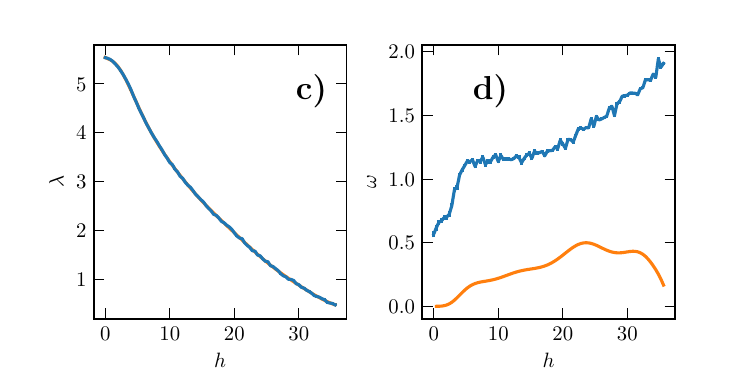}
\caption{Dependece of $\lambda$ and $\omega$ on the applied field $h$, with $\eta$ (subplots a and b) and $\eta_{1}$ (subplots c and d). Each data point is calculated from 64 simulations of $L=512$ and $T=256$. Error bars come from the same procedure as the error bars in Table \ref{tab:params}. Orange lines in subplots b) and d) are $c (h)$, and come from fitting an 8th order polynomial to subplots a) and c) then extracting the slope $d\lambda/dh$ from the polynomial. [I think the orange lines are a little messed up. Also: I'm listing $\omega$ as the standard deviation everywhere else, but here its variance.] }
\label{fig:hdependence}
\end{figure}

Having determined that the magnetization is indeed roughly diffusive, we examined the coupling of the maximal Lyapunov exponent $\lambda$ to the applied field $h$. For a given field, $\lambda$ is calculated by running many simulations, calculating the mean of the log of the perturbation at each time, and averaging over all simulations. Then a line is fitted to a plot of the log of the perturbation versus time, and the fitted slope is $\lambda$. Figure \ref{fig:hdependence}a shows the coupling of $\lambda$ to the applied field $h$. As anticipated, $\lambda$ gets smaller as the system becomes more strongly magnetized. 

In the DPRM, switching on energy conservation increased the HPW fluctuation index $\omega$ by creating regions of high and low energy. In the spin chain, we want to accomplish the same thing. Figure \ref{fig:hdependence}a shows that the Lyapunov exponent is coupled to the magnetization via some highly nonlinear function. In order for small local fluctuations in the magnetization to have the strongest effect, we want to put the system at the point where the coupling of $\lambda$ to the magnetization $m$ is strongest, so that small fluctuations have the largest impact. Specifically, we believe that the coupling will be related to the quantity
\begin{equation}\label{eq:13}
c (h) = \left (\frac{\partial \lambda}{\partial m} \sigma_{m} \right),
\end{equation}
where $m$ is the average magnetization and $\sigma_{m}$ is the rms size of fluctuations in the magnetization.

From Figure \ref{fig:hdependence}a, we can extract the slope $d\lambda/dh$ by fitting a polynomial and taking the derivative. To calculate $c (h)$ from $d\lambda/dh$, we observe that 
\begin{equation}\label{eq:14}
\frac{\partial\lambda}{\partial m} = \frac{\partial \lambda}{\partial h } \frac{d h }{dm}.
\end{equation}
The quantity $dh/dm$ is easily calculated by noting that 
\begin{equation}\label{eq:15}
m = \frac{1}{Z}\int_{-1}^{1}ze^{hz}dz = \frac{\partial \log Z}{\partial h}
\end{equation}
To find $\sigma_{m}$, we use a tool from statistical mechanics, 
\begin{equation}\label{eq:16}
\frac{\partial^{2}\log Z}{\partial h^{2}} = \sigma_{m}^{2}.
\end{equation}
But in fact, this tells us that $ \sigma_{m}^{2} = dm/dh$. So 
\begin{equation}\label{eq:17}
c (h) = \frac{\partial \lambda}{\partial h} \frac{dh}{dm} \sqrt{\frac{dm}{dh}} = \frac{\partial \lambda}{\partial h} \sqrt{\frac{dh}{dm}} = \frac{\partial\lambda}{\partial h} \left (\frac{1}{h^{2}}-\csch^{2}h\right)^{-1/2}.
\end{equation}

\begin{figure}[b!]
\centering
\includegraphics{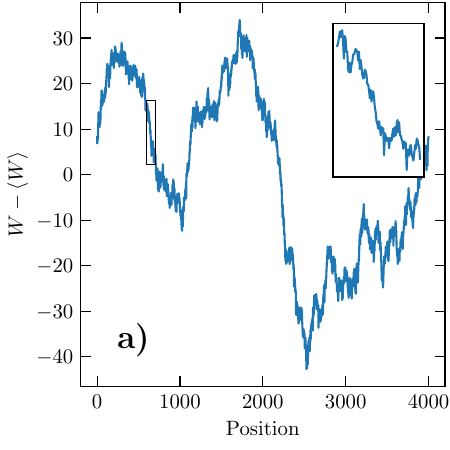}\hfill
\includegraphics{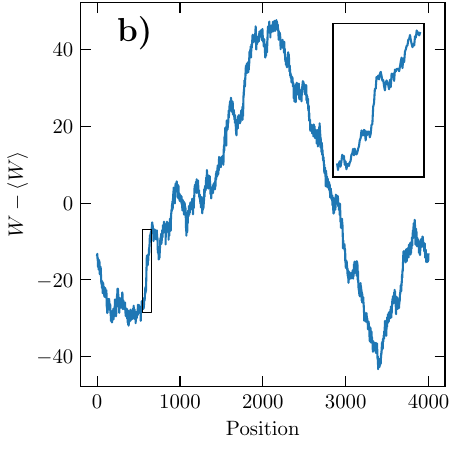}
\includegraphics{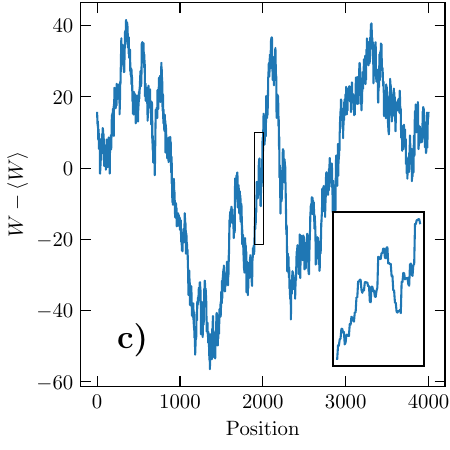}\hfill
\includegraphics{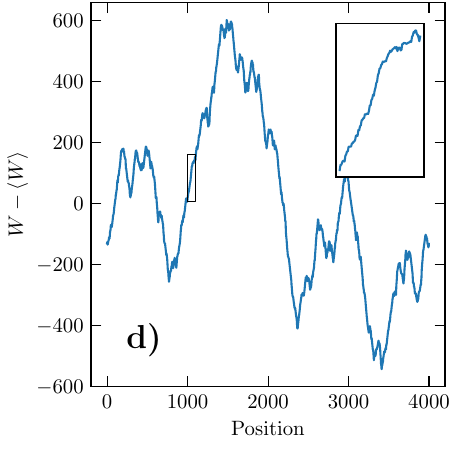}
\caption{Plots of $W=\log (|P|)$ vs. $x$ in the spin chain model. The top row uses $\eta$, the bottom row uses $\eta_{1}$. Subplots a) and c) have $h=0$, subplot b) has $h=3.5$, subplot d) has $h=8$. Insets show the contents of the small boxed regions of the plots. For all subplots, $T=1024$ and $L=4096$.  }
\label{fig:scalefreechain}
\end{figure}

The fluctuation index $\omega$ is here defined as the scaling power of the standard deviation of the log of the perturbation. Calculating $\omega$ as a function of $h$ results in Figure \ref{fig:hdependence}b. Also included in this plot is an orange line representing $c (h)$, as determined from the slope of Figure \ref{fig:hdependence}a. The two lines have very similar shapes, as predicted. Both are maximal near $h=3.5$, so this value of $h$ should have behavior most similar to the DPRM with energy conservation. However, as Figure \ref{fig:scalefreechain}b shows, the other qualitative markers are absent. In the DPRM, energy conservation was marked by a shift from rough, scale-free spatial path-weight profiles, to smooth triangular path-weight profiles. Both subplots a) and b) of Figure \ref{fig:scalefreechain} (with $h=0$ and $h=3.5$) look rough and scale-free. There appears to have been no phase transition like in the DPRM.

At this point, we return to Figure \ref{fig:diffusion}b. Earlier, we observed that this behavior was messy and not very diffusive looking. We believe that this messiness might be partly to blame for the failure of the spin chain to undergo a transition similar to the DPRM. We therefore modified the model in an attempt to clean up Figure \ref{fig:diffusion}b. Each spin had been precessing by an angle $\theta$ determined by the original function 
\begin{equation}\label{eq:18}
\theta = \eta \left(|\vec{S} (x,t)+\vec{S} (x\pm 1,t)|\right)
\end{equation}
with $\eta (x)=\pi x/4$. We decided to replace $\eta (x)$ with 
\begin{equation}\label{eq:19}
\eta_{1} (x) = \frac{\pi}{16} x (x^{2}-4)^{2}.
\end{equation}
This new function was chosen to have the feature that when the spins were perfectly aligned $(x=2)$ it went to 0. This is designed to suppress the waves visible in Figure \ref{fig:diffusion}b. $\eta_{1}$ was also chosen to have a maximum rotation angle of $\pi/2$, same as $\eta$. It is also nonlinear, which we believed would make the chaos stronger.

The behavior of the system with $\eta_{1}$ was overall much more like the DPRM model. Figure \ref{fig:diffusion}c and d show that the $z$ component of the spin moves roughly diffusively at all magnetizations, an improvement over $\eta$. Figure \ref{fig:hdependence}c shows that the Lyapunov exponent, and thus the chaos, are nearly an order of magnitude stronger than before, which will hopefully make the effects more visible. 

Figure \ref{fig:hdependence}d shows what looks like three separate regions. For $h<5$, the energy fluctuation index $\omega$ is increasing as the coupling between the magnetization and the Lyapunov exponent is turned on. For $5<h<20$, there is a fairly stable area with $\omega \sim 0.57$ where the coupling has been turned on. Then as $h>20$, some other effect seems to take over. We believe that at this point, most of the spins are entirely immobile and the chaos doesn't develop at all, except for a few streams where it grows linearly in time. This would results in energy fluctuations that grow linearly, consistent with the apparent asymptotic limit $\omega \to 1$.

Further evidence that something qualitatively different is happening in the region $5<h<20$ can be seen in Figure \ref{fig:scalefreechain}d, which shows the spatial profile of the HPW with $\eta_{1}$ and $h=8$. Unlike all the other subplots in Figure \ref{fig:scalefreechain}, this subplot has the kind of smooth triangular profile characteristic of Figure \ref{fig:scalefreedpm}b and c.

\begin{figure}[ht]
\centering
\includegraphics{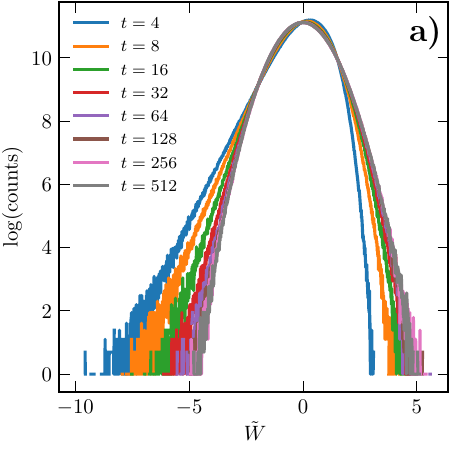}
\includegraphics{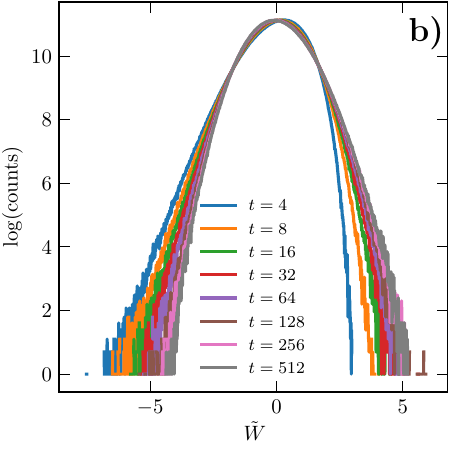}
\caption{Distributions of ``\HWPWs[]'' $W (x,t) = \log (P (x,t))$ in the spin chain model. Subplot a) shows $\eta$ and $h=3.5$, which is the $h$ value that produced the highest value of $\omega$. Subplot b) shows $\eta_{1}$ and $h=8$. $\tilde W$ is rescaled to have zero mean and unit variance, so these plots show only the shape and skew of the distributions. Distributions are accumulated from 256 simulations with length $L=1024$.} 
\label{fig:wdistchain}
\end{figure}

A final piece of evidence is Figure \ref{fig:wdistchain}, which shows the distributions of $W (x,t)$. These distributions both start skewed the opposite direction from the distributions in Figure \ref{fig:wdistdpm}. However, the distribution in subplot d eventually switches and ends up skewed the same direction as Figure \ref{fig:wdistdpm}, while the distributions in subplot b (and also all the other situations) never end up skewed the other direction. 

\section{Conclusion}

In this paper, we have shown that the introduction of a diffusive conserved quantity can have substantial effects on classical chaos models. In the DPRM, adding diffusive transport to the energy landscape increased the \HWPW fluctuation index $\omega$. It also dramatically changed the \HWPW spatial profile, at long times eliminating the characteristic roughness of Figure \ref{fig:scalefreedpm}a in favor of large, smooth, triangular features like those in Figure \ref{fig:scalefreedpm}b. We found that this was because the HWPs (Figure \ref{fig:paths}) were forming long stems with slow wandering, which branched ballistically at late times to form the characteristic triangles of Figure \ref{fig:scalefreedpm}. In this new DPRM, and even more so in the dipole-conserved DPRM, the physics were primarily controlled by the locations and sizes of ``streams'' of high $z$. 

In our discrete-time implementation of the Heisenberg spin chain, the magnetization is conserved and exhibits diffusive transport. Moreover, there is a strong coupling between the magnetization and the Lyapunov exponent. Therefore, local fluctuations in the magnetization lead to locally different Lyapunov exponents, which spread diffusively like our diffusive DPRM energy landscape. Using a slightly modified precessional dynamics ($\eta_{1}$), we were able to move the system into the same regime as the DPRM with energy conservation. In particular, we reproduced the smooth triangles of the spatial profile of the perturbation (Figure \ref{fig:scalefreechain}), and the increase in the fluctuation index $\omega$. This transition indicates that the development of the chaos in this spin chain is primarily controlled by fluctuations in the local magnetization, likely developing most strongly along ``streams'' of relatively low magnetization (in analogy to the streams of high $z$ in the DPRM). The problem of understanding the development of the chaos in this spin chain thus becomes, to some extent, an issue of understanding the hydrodynamics of the magnetization in the spin chain.

These results bear some relationship to the conclusions of \cite{pikovsky1998dynamic} and \cite{pazo2016diverging}. In \cite{pikovsky1998dynamic}, the authors found that nontrivial long-range correlations produced behavior different from that predicted by an analogy to the KPZ equation. In our original DPRM, the characteristic roughness of Figure \ref{fig:scalefreedpm}a is predicted by the KPZ equation; so the elimination of this roughness in Figure \ref{fig:scalefreedpm}b is in fact evidence that the diffusive transport of the energy breaks the analogy to the KPZ equation in a related way.
In \cite{pikovsky1998dynamic}, the authors found that long-range correlations resulting from the hydrodynamics of the system were responsible for diverging fluctuations of the finite-time Lyapunov exponents, and predict that this effect might occur as well in a situation much like ours.

From here, this project could be taken in many different directions. The separation of regimes in the DPRM with subdiffusive transport shown in Figure \ref{fig:singleorigin} is a very interesting effect. Since fracton models share many of the qualities that produced this behavior (conservation of both a quantity and its dipole moment), one next step could be trying to study this chaos in such a fracton model. Another next step could be studying the different regimes in Figure \ref{fig:hdependence}d more carefully. In particular, the \HWPW fluctuation index goes to $1$ at high $h$, which might bear some relation to the diverging fluctuations of the Lyapunov exponent demonstrated in \cite{pazo2016diverging}. A third direction could be to study these effects in a continuous-time spin chain model. This would be more complicated, as energy and magnetization would both be conserved, but it would also be a more physically realistic system.

\newpage
\clearpage
\addcontentsline{toc}{section}{References}
\bibliographystyle{h-physrev}
\bibliography{jp2}


\newpage
\clearpage

\section{Supplementary materials}

\subsection{Verifying transport in DPRM models with conservation}

\subsection{Propagation of the perturbation}\label{sec:prop-pert}

We derive here the propagation of the perturbation to first order in $\epsilon$ as described in Section \ref{sec:class-heis-spin}. Recall that in each pairwise spin-spin interaction between a spin $\vec{S}_{1}=\vec{S} (x,t)$ and another spin $\vec{S}_{2}=\vec{S} (x+1,t)$, the first spin $\vec{S}_{1}$ is rotated by an angle $\theta = \eta (|\vec{S}_{1}+\vec{S}_{2}|)$ about the axis of their vector sum,
\begin{equation}\label{eq:23}
\vec{n} = \frac{\vec{S}_{1}+\vec{S}_{2}}{|\vec{S}_{1}+\vec{S}_{2}|}.
\end{equation}
This rotation can be implemented conveniently using quaternion multiplication. We write $\vec{S}_{1} = S_{1,x}i+S_{1,y}j+S_{1,z}k$, and likewise for $\vec{S}_{2}$. The rotation of $\vec{S}_{1}$ can then be performed by defining 
\begin{equation}\label{eq:25}
\vec{q} = e^{\frac{\theta}{2} \vec{n}} = \cos\frac{\theta}{2} + \sin\frac{\theta}{2}\vec{n}
\end{equation}
and taking
\begin{equation}\label{eq:24}
\vec{S} (x,t+\frac{1}{2}) = \vec{q} \vec{S}_{1}\vec{q}^{-1},
\end{equation}
where 
\begin{equation}
\vec{q} = e^{-\frac{\theta}{2} \vec{n}} = \cos\frac{\theta}{2} - \sin\frac{\theta}{2}\vec{n}.
\end{equation}
Recall that we are considering a perturbed state $\vec{S}' (x,t) = \vec{S} (x,t)+\epsilon \vec{P} (x,t)$. To propagate the perturbation, we thus need to calculate 
\begin{equation}\label{eq:26}
\frac{d \vec{S}' (x,t+\frac{1}{2})}{d\epsilon}\biggr\rvert_{\epsilon= 0} = \vec{P} (x,t+\frac{1}{2}).
\end{equation}
This derivative is fairly straightforward. In the following, we will drop the $\rvert_{\epsilon = 0}$ for simplicity of notation.
\begin{align}
\frac{d \vec{S}' (x,t+\frac{1}{2})}{d\epsilon}\biggr\rvert_{\epsilon= 0} =& \frac{d}{d\epsilon}\left[    
\vec{q} \vec{S}_{1}'\vec{q}^{-1}
 \right] \\
=& \frac{d\vec{q}}{d\epsilon} \vec{S}_{1}\vec{q}^{-1} + \vec{q}\frac{d\vec{S}_{1}'}{d\epsilon} \vec{q}^{-1} + \vec{q}\vec{S}_{1} \frac{d\vec{q}^{-1}}{d\epsilon} \\
=& \left ( -\frac{1}{2}\frac{d\theta}{d\epsilon}\sin \frac{\theta}{2} 
+ \frac{1}{2}\frac{d\theta}{d\epsilon}\cos\frac{\theta}{2} \vec{n} + \sin\frac{\theta}{2} \frac{d\vec{n}}{d\epsilon}
\right) \vec{S}_{1} \vec{q}^{-1} \nonumber \\
&+ \vec{q} \vec{P}_{1} \vec{q}^{-1} \nonumber \\
&+ \vec{q}\vec{S}_{1}
\left ( -\frac{1}{2}\frac{d\theta}{d\epsilon}\sin \frac{\theta}{2} 
- \frac{1}{2}\frac{d\theta}{d\epsilon}\cos\frac{\theta}{2} \vec{n} - \sin\frac{\theta}{2} \frac{d\vec{n}}{d\epsilon}
\right) .\label{eq:28}
\end{align}
There are two quantities in this expression which need calculating, $\frac{d\theta}{d\epsilon}$ and $\frac{d\vec{n}}{d\epsilon}$.
These are given by
\begin{equation}
\frac{d\theta}{d\epsilon} = \eta' (|\vec{S}_{1}+\vec{S}_{2}|) \frac{d}{d\epsilon}|\vec{S}_{1}'+\vec{S}_{2}'|
\end{equation}
and
\begin{equation}
\frac{d\vec{n}}{d\epsilon} = \frac{(\vec{P}_{1}+\vec{P}_{2})|\vec{S}_{1}+\vec{S}_{2}| - (\vec{S}_{1}+\vec{S}_{2})\frac{d}{d\epsilon}|\vec{S}_{1}'+\vec{S}_{2}'|}{|\vec{S}_{1}+\vec{S}_{2}|^{2}},
\end{equation}
where 
\begin{align}
\frac{d}{d\epsilon}|\vec{S}_{1}'+\vec{S}_{2}'| =& \frac{d}{d\epsilon} \sqrt{(\vec{S}_{1}'+\vec{S}_{2}')\cdot (\vec{S}_{1}'+\vec{S}_{2}')}  \\
=& \frac{\vec{P}_{1}\cdot \vec{S}_{2}+\vec{P}_{2}\cdot\vec{S}_{1}}{|\vec{S}_{1}+\vec{S}_{2}|},
\end{align}
and in the final line we have used the fact that $\vec{S}_{a}\cdot\vec{P}_{a}=0$. Plugging all of these results back into Equation \eqref{eq:28} gives $\vec{P} (x,t+\frac{1}{2})$, the perturbation on site $1$ after one round of pairwise spin-spin interactions.

\end{document}

%% file: liouville.tex
\begin{tikzpicture}[scale=0.7]
\draw  [] (0,3.8) to [out=0,in=90] (3.8,0);
\draw  [] (0,5) to [out=0,in=90] (5,0);
\draw  [] (0,3.8) to  (0,5);
\draw  [] (3.8,0) to  (5,0);

\draw  [<->] (1,4.4) to [out=-15,in=105] (4.4,1);

\draw (-1,-1) to (-1,1) to (1,1) to (1, -1) to (-1, -1);

\draw [->, double] (1.2,1.2) to (2.5,2.5);

\draw [<->] (0,-0.4) to (0,0.4);

\node at (0.3, 0) {\small $\epsilon$};

\node at (3, 3) [rotate=-45] {\small $\epsilon e^{\lambda t}$};

\end{tikzpicture}

%% file: interactionscheme.tex
\newcommand{\spina}[3]{
\draw (#1, #2) circle [radius=0.22];
\node at (#1, #2-0.4) {\scriptsize #3};
}
\newcommand{\spinb}[3]{
\draw (#1, #2) circle [radius=0.22];
\node at (#1, #2+0.4) {\scriptsize #3};
}
\newcommand{\spin}[2]{
\draw (#1, #2) circle [radius=0.22];
}
\newcommand{\interact}[2]{
\draw (#1-0.2, #2-0.2) rectangle (#1+0.2, #2+0.2);
}
\newcommand{\upright}[2]{
\draw [->] (#1+0.25,#2+0.25) -- (#1+0.75,#2+0.75);
}
\newcommand{\upleft}[2]{
\draw [->] (#1-0.25,#2+0.25) -- (#1-0.75,#2+0.75);
}
\begin{tikzpicture}[node distance = 2cm, auto, scale=1]
\spina{1}{0}{$\vec{S} (x-1,t)$}
\spina{3}{0}{$\vec{S} (x,t)$}
\spina{5}{0}{$\vec{S} (x+1,t)$}
\spina{7}{0}{$\vec{S} (x+2,t)$}
\spin{1}{2}
\spin{3}{2}
\spin{5}{2}
\spin{7}{2}
\spinb{1}{4}{$\vec{S} (x-1,t+1)$}
\spinb{3}{4}{$\vec{S} (x,t+1)$}
\spinb{5}{4}{$\vec{S} (x+1,t+1)$}
\spinb{7}{4}{$\vec{S} (x+2,t+1)$}

\interact{2}{1}
\interact{6}{1}
\interact{0}{3}
\interact{4}{3}
\interact{8}{3}

\upright{1}{0}
\upright{2}{1}
\upright{3}{2}
\upright{4}{3}
\upleft{3}{0}
\upleft{2}{1}
\upleft{1}{2}
\upright{0}{3}

\upright{5}{0}
\upright{6}{1}
\upright{7}{2}
\upleft{8}{3}
\upleft{7}{0}
\upleft{6}{1}
\upleft{5}{2}
\upleft{4}{3}

\upright{-1}{2}
\upleft{0}{3}
\upleft{9}{2}
\upright{8}{3}
\end{tikzpicture}